\documentclass[aps,reprint,superscriptaddress,longbibliography,nofootinbib, pra]{revtex4-1}

\usepackage{amsmath, amssymb, mathtools, physics}

\usepackage{csquotes}

\usepackage{graphicx}

\usepackage[colorlinks=true, allcolors=monbleu]{hyperref}

\usepackage{xcolor}
\definecolor{monbleu}{RGB}{76, 114, 176}

\graphicspath{{figures/}}

\begin{document}
\title{Wasserstein recurrence networks for multiscale time series pattern analysis}
\date{\today}
\author{B\'eatrice D\'esy}
\email{desybeatrice@gmail.com}
\affiliation{School of Information Management, Victoria University of Wellington, Wellington, 6140, New Zealand}
\affiliation{Antarctic Research Centre, Victoria University of Wellington, Wellington, 6140, New Zealand}

\author{Nicholas R. Golledge}
\affiliation{Antarctic Research Centre, Victoria University of Wellington, Wellington, 6140, New Zealand}
\affiliation{Te Pūnaha Matatini, Auckland 1142, New Zealand}

\author{Hana Ishii}
\affiliation{Antarctic Research Centre, Victoria University of Wellington, Wellington, 6140, New Zealand}

\author{Markus Luczak-Roesch}
\affiliation{School of Information Management, Victoria University of Wellington, Wellington, 6140, New Zealand}
\affiliation{Te Pūnaha Matatini, Auckland 1142, New Zealand}

\begin{abstract}
Time series data are often generated by systems which operate on multiple temporal scales, of which Earth's climate system is a paramount example. Variations in global climate are recorded in paleo-environmental archives as temporal patterns across a wide range of time scales, from seasonal or decadal to multi-millennial. In this context, recurrence analysis, where repeating patterns are identified in time series, is limited by the underlying properties of the distance function used and of the time series data themselves, especially in terms of temporal resolution and scale dependence. In this paper, we present a novel recurrence analysis framework designed for multiscale time series data with abrupt changes and irregular temporal resolution as found in paleoclimate records. We introduce a simple mathematical transform to use the $1-$Wasserstein distance for recurring pattern detection in time series. The scale invariance of $1-$Wasserstein distance distributions between patterns in Brownian motion is demonstrated numerically, which provides a principled threshold choice for recurrences. At any time scale, recurrences are defined as local minima of the distance, granted that they are below a threshold given by the probability of encountering patterns at least as similar in one-dimensional Brownian motion. Recurrences can be further combined according to a non-overlapping condition to yield a distinct set of multiscale recurring events. We provide examples of climatic applications from ice-rafted debris and ice core records, where detected recurrences have durations spanning over two orders of magnitude. Our method extends recurrence analysis to more complex time series data and provides new avenues for statistical identification and analyses of recurring events at multiple temporal scales. 
\end{abstract}

\maketitle


\section{Introduction}
At the root of most knowledge systems that describe and predict natural phenomena, like physical laws, indigenous science, and artificial intelligence, lies the ability to detect and store information about patterns that recur over time. When repeating patterns are known, it is possible to start exploring the underlying processes that generated them. This is a challenging task when studying the Earth's climate system, which operates on many different temporal and geographical scales all at once. Paleoclimate archives like ice cores and marine sediment cores capture rich data on both local and global climate patterns, but the data can be noisy, sparse, and more often than not unevenly sampled in time. Yet, it is important to study past climates in order to gain a wider understanding of possible trajectories and tipping points in the context of anthropogenic climate change~\cite{Harrison2025_Paleoclimate}.

Methods to investigate paleoclimate data mostly fall into two main categories: directly analyzing the time series or using the data to calibrate climate models~\cite{Haywood2019_What, Lunt2024_Paleoclimate}. When bridging climate models and proxy data, the ubiquitous trade-off between model complexity and overfitting arises. Simpler models provide insight into physical processes but cannot exactly recreate empirical records. Complex models can reproduce features of elaborate paleoclimate data sets~\cite{Tierney2025_Advances}, but those models are very computationally costly to run and may be highly parameterized~\cite{Haywood2019_What}. To look into paleoclimatic data directly, existing literature focuses on frequency analysis (\emph{i.e.} Fourier, wavelets, power spectrum)~\cite{Weedon2022_Problems}, which requires time series preprocessing and creates non-trivial dependencies with the age model~\cite{Meyers2019_Cyclostratigraphy}. Methods that require uniform temporal sampling make an implicit assumption about what is noise -- short temporal scale, abrupt changes -- but in some climate records, abrupt changes capture real climate processes. On the other hand, many methods require detrending, which also erases important features in the data. Inspecting specific epochs or moments in time has provided numerous insights on past events like glacial terminations~\cite{Barker2025_Distinct}, but is inherently restricted. Here we present a novel method that offers an avenue to intuitively explore all the richness of raw temporal data whilst making minimal assumptions, simply asking what happens again and again, at all accessible time scales, without assuming an underlying model. 

Recurrence analysis has proven successful in providing insights into various dynamical systems, including climate~\cite{Westerhold2020_astronomically}, heart rates and diseases~\cite{Yang2019_New}, and the human brain~\cite{Varley2022_Network}, as well as detecting dynamical transitions in non stationary time series~\cite{Donges2011_Identification, Marwan2009_Complex}. The core idea of recurrence analysis and time-delay embeddings is to identify values or patterns that recur over time~\cite{Eckmann1987_Recurrence, Marwan2007_Recurrence}. This family of approaches has been used to identify states in different elements of the global climate system, for instance to quantify transitions between epochs of climate regularity and periods of abrupt changes in precipitation data~\cite{Zhang2016_Application, Agarwal2022_complex} and palaeoclimate proxy records~\cite{Marwan2009_Complex, Donges2011_Identification, Westerhold2020_astronomically}, relation between those transitions and human evolution~\cite{Donges2011_Nonlinear, Trauth2019_Classifying}, and more recently exploring the potential for flood prediction~\cite{Zhang2026_High}. In practice, existing methods are challenging to use due to requiring data preprocessing and non-trivial parameter choices~\cite{Marwan2011_How}, like which distance function to use~\cite{Donner2010_Ambiguities} and the similarity threshold to define a recurrence~\cite{Schinkel2008_Selection, Kraemer2018_Recurrence}. For instance, to detect recurrences at a comparable time scale on paleoclimate data with uneven temporal resolution, one either has to interpolate~\cite{Westerhold2020_astronomically}, adapt the number of data points used depending on the resolution~\cite{Donges2011_Nonlinear}, or use non-standard metrics~\cite{Ozken2018_Recurrence}. Using Euclidean time-delay embeddings for non-stationary time series creates a non-trivial interplay between the choice of threshold and the embedding dimension (effectively the temporal duration of recurring patterns)~\cite{Kraemer2018_Recurrence}. A common approach is to set a fixed recurrence rate~\cite{Marwan2009_Complex}, or proportion of recurrences detected, which is independent from the number of recurring events in the underlying dynamical system. Local minima-based methods are an interesting avenue to address the threshold choice problem~\cite{Schultz2011_Local}, which we develop in this work. Another limitation of existing methods is the study of recurrences of different durations. The well-known "curse of dimensionality" for $L^p$ metrics (which include the standard Euclidean distance) makes it challenging to study events at multiple time scales all at once~\cite{Varley2022_Network}, especially across orders of magnitude in event duration. In the climate system, the same short-scale event with different underlying conditions could yield a completely different long-term response~\cite{Hanna2024_Short}. 

Here we provide a new method to detect recurrences in multiscale time series and an analysis framework to map those to a network of recurring patterns over time. Our recurrence detection method is designed for realistic (noisy, sparse, unevenly sampled) temporal data that operates on many different time scales all at once. Our framework builds on the assumption that physical processes generating time series observations reveal themselves through recurring events over time. All of the data is used, regardless of the temporal sampling, which facilitates detection of recurring events across different records. For multiscale time series, we introduce a choice of threshold with an intuitive interpretation in terms of confidence level. The main outputs of this work are twofold: i) a universal method to detect recurring patterns in messy time series data and ii) a distinct set of recurring events, spanning multiple time scales, on which statistical analyses can be performed. 

This work is structured as follows. First, we introduce some mathematical notation and how to use the well-known 1-Wasserstein distance for time series pattern recurrence detection. In \textsection~\ref{sec:brownian}, we study the distribution of Wasserstein distances between patterns in Brownian motion and show numerical invariance, which provides a principled threshold choice for pattern recurrences across multiple time scales. Building on those two sections, recurrences are defined as statistically significant closest patterns. We follow with a description of the multiscale network mapping framework in \textsection~\ref{sec:network}. Finally, two applications on paleoclimate records are presented to illustrate our method in \textsection~\ref{sec:palaeo_results}, on both very abrupt ice rafted-debris data for which Brownian motion is not an adequate null model, and an ice-core global temperature proxy over multiple ice ages to demonstrate the recurrence network across time scales.\\

The applications we introduce are paleoclimatic, but our framework can be used to analyze any type of temporal data for which existing methods present similar drawbacks. Different facets of the methods can be used independently. In particular, one does not need to delve into network time series analysis to use the Wasserstein distance for repeated pattern detection in temporal data, which is exemplified in \textsection\ref{sec:ird}. Our mapping from time series to a network of recurring moments in time is inspired by prior work to study text data~\cite{Luczak-Roesch2015_When, Luczak-Roesch2018_What}. Predictive applications are outside the scope of this paper, and so is the question of causal links between recurrences.

%

\section{\label{sec:recurrences} Wasserstein distance for recurrence analysis}

Here we define recurring patterns in time series as local minima of the 1-Wasserstein distance. Our framework to use the 1-Wasserstein metric is well suited for unevenly sampled time series data that operate over many different time scales all at once, as most other metrics would require interpolation or sub-sampling and re-parameterization for treating different recurrence durations. A local minimum is the closest definition of an exact recurrence that can be extended to imperfect quantitative data. This makes our method particularly fit for realistic time series and detecting similar patterns in different time series. An additional thresholding method to identify significant local minima in multiscale time series will be presented in the following \textsection~\ref{sec:brownian}. \\

Let \(Y=\{(y_i, t_i)\}_{i=1,\hdots,m}\subset\mathbb{R}^2\) be a real time series, \emph{i.e.} a discrete ensemble of pairs \((y_i, t_i)\), where \(y_i\) are \emph{values} and \(t_i\) are their \emph{time stamps} \(t_0<t_1<\hdots<t_{m-1}<t_m\), which provide a total order on \(Y\). An \emph{event} or \emph{pattern} \(x\) in the time series is defined as a consecutive subset of \(Y\) covering a given time interval $[t, t+\ell\tau]$. It is identified as a function of its starting time \(t\) and its duration \(\ell\tau\),
\begin{equation}
\label{eq:pattern}
x(t, \ell\tau) = \{(y_i, t_i)\in Y\,|\, t\leq t_i\leq t+\ell\tau\},
\end{equation}
with \(\ell\in\mathbb{N}\) its duration in discrete time increments \({\tau\in\mathbb{R}}\). In this work, we use "time scale" as a description of event length or duration \(\ell\), thus "multiple time scales" refers to events of various durations. Patterns are equivalent to information tokens (like words or hashtags) in symbolic data, since they provide coarse-grained and temporally ordered information about the evolution of the system that generated the time series. A sequence of zeroes can be defined as an acceptable event or not depending on the context (e.g., for ice-rafted debris data, a sequence of zeroes is an absence of data, whereas for temperature data, it is a valid pattern).\\ 

The Wasserstein\footnote{This distance is also called Earth mover's distance in computer science and Kantorovich–Rubinstein distance, after scholars whose work preceded Wasserstein's.} metric is a robust and elegant way to quantify the similarity between probability distribution functions (pdf) on metric spaces~\cite{Panaretos2019_Statisticala, Villani2009_Optimal}.  Intuitively, it is the least work that would be required to \enquote{transform} one pdf into the other, where work is proportional to the probabilistic \enquote{mass} (area under the curve) and how much distance it has to be transported as quantified by the metric. It quantifies pattern similarity in a way that is more aligned with perceptual similarity in computer vision~\cite{Rubner2000_Earth, Crook2024_linear} and has been used for time series classification~\cite{Muskulus2011_Wasserstein}, analysis~\cite{Cazelles2021_WassersteinFourier}, and numerous machine learning applications~\cite{Gabriel2019_Computational, Kolouri2017_Optimal}. Given two probability measures \(\mu\) and \(\nu\) over a metric space \((\mathcal{X}, d)\) equipped with the distance function \(d\), and \(p\in[1,\infty)\), the \(p\)-Wasserstein metric is defined as
\begin{equation} \label{eq:wasserstein}
W_p(\mu, \nu) = \bigg(\inf_{\pi\in\Pi(\mu, \nu)} \int_{\mathcal{X}\times\mathcal{X}} \big[d(x,y)\big]^p\dd\pi(x,y)\bigg)^{1/p},
\end{equation}
where the infimum is taken over all possible joint probability distributions \(\pi\) having both \(\mu\) and \(\nu\) as marginals. Exponent values \(p=1\) and \(p=2\) are most useful, with the latter particularly suited when there is more structure in the underlying metric space~\cite{Villani2009_Optimal}. Since here it is the one-dimensional real line, we only consider \(p=1\) in what follows. With \(d(x,y)=|x-y|\), \(x,y\in\mathbb{R}\), an alternative and computationally efficient formulation of \(W_1\) is
\begin{equation} \label{eq:wasserstein_cdf}
W_1(\mu, \nu) =\int_\mathbb{R} \Big|F^{-1}_\mu(t)-F^{-1}_\nu(t)\Big|\,\dd t,
\end{equation}
given in terms of the cumulative distribution functions (cdf) \(F_\mu\) and \(F_\nu\) and their reciprocals \({F^{-1}_i(F(x)) = \min\{x\,|\, F^{-1}_i(F(x))=x\}}\), \(i=\mu, \nu\). This result is attributed to Dobrushin~\cite{Dobrushin1970_Prescribing} and an alternate derivation is presented in Ref.~\cite[Appendix]{Ramdas2015_Wasserstein}. With this formulation, the optimal transport problem is a linear programming one~\cite{Kolouri2017_Optimal}.

In order to use the Wasserstein metric in signal analysis, subsequences of time series have to be transformed into probability measures. Consider a nonempty pattern \(x(t, \ell\tau)=\{(y_i, t_i)\}_{i=1,\hdots,n_x}\subset Y\). Following the notation from Ref.~\cite{Muskulus2011_Wasserstein}, let us define a mapping from \(x\) to a discrete measure \(\mu(x)\) over the interval \([0,1]\) as the weighted sum 
\begin{equation}\label{eq:discrete_measure}
\mu(x) = \sum_{i=1}^{n_x} \alpha_i\, \delta\left(\frac{t_i-t}{\ell\tau}\right),
\end{equation}
where \(\delta\) is the Dirac delta function, \(n_x=|x|\) is the number of data points in \(x\) and weights \(\alpha_i\) are normalized values of the time series such that,
\begin{equation}\label{eq:normalisation}
\alpha_i = \frac{y_i-\min\{y_i\}+f(x)}{\sum_{i=1}^{n_x} (y_i-\min\{y_i\}+f(x))}.
\end{equation}
The \(f(x)\) term is a lower bound defined as
\begin{equation}\label{eq:lower_bound}
f(x) = \begin{cases} 1/n_x & \text{if } \min\{y_i\}=\max\{y_i\},\\
\frac{|\max\{y_i\}- \min\{y_i\}|}{n_x} & \text{otherwise.}
\end{cases}
\end{equation}
The lower bound on the translation of values depends on their range and on the number of points \(n_x\) in the subsequence. If the range is bigger than the number of points, then even the lowest value still has a lot of weight, whereas if the number of points is very high, the lowest value gets closer to zero. This transformation from a time series pattern to a discrete probability measure is merely a rescaling in time and values; therefore it preserves the shape of the pattern. 
\begin{figure}
\includegraphics[width=\columnwidth]{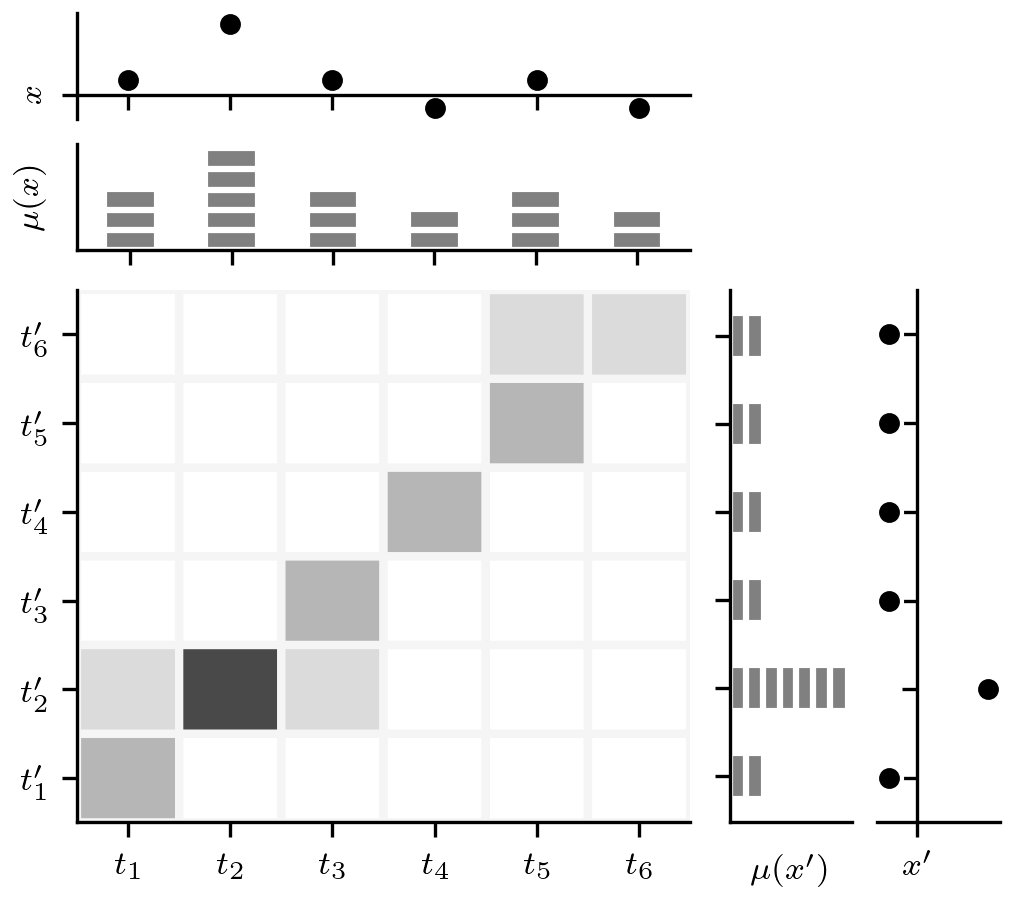}
\caption{\label{fig:wasserstein} Principle of the Wasserstein distance between discrete time series patterns \(x\) and \(x'\). Both original patterns \(x\) and \(x'\) are shown by black dots (top and right), next to their transformation to discrete measures \(\mu(x)\) and \(\mu(x')\), represented as histograms. The matrix represents a joint probability distribution \(\pi\) that has \(\mu(x)\) and \(\mu(x')\) as marginals, with probability density proportional to darkness. }
\end{figure}

The 1-Wasserstein distance between \(x\) and any other pattern \(x'\) in the time series (or another time series) can be computed as a composition of \(\mu\) with \(W_1\),
\begin{equation}\label{eq:my_wasserstein}
W(x,x') := W_1\big(\mu(x), \mu(x')\big),
\end{equation}
with \(W_1\) computed with Eq.~\eqref{eq:wasserstein_cdf}. This process is illustrated in Fig.~\ref{fig:wasserstein} for a single pattern pair.
\(W\) above is maximal when the distance is computed between patterns that are single data points at each end of the time interval. Its value is then \(1\), which means \(W\) is conveniently bounded between 0 and 1. The rescaling in time could be equivalently done after measuring Wasserstein distances according to the property \(W(aX, aY) = |a|W(X,Y)\) \cite[\textsection 2]{Panaretos2019_Statisticala} with \(a=\ell\tau\). For 1D distributions, the numerical implementation of the Wasserstein distance reduces to a discrete numerical integral based on cumulative distribution functions. The function \texttt{wasserstein\_distance} provided in the SciPy package is used in our Python code and examples. 

The Wasserstein distance is particularly interesting for comparing subsequences of time series since it encompasses distance in both time, the underlying metric space, and the shape of the signal. Unlike information-theoretic divergences, it does not depend on the size of the pattern and does not require patterns to contain the same number of data points to be compared. The Wasserstein distance as defined above is able to capture similarity between time series patterns at many different time scales and does not require preprocessing, even when the temporal sampling rate varies. This is particularly relevant when processes at multiple time scales are interwoven, because performing interpolation is making an \emph{ad hoc} choice about what is noise and what is signal. Another common assumption when interpolating is that any abrupt change is attributed to noise, whereas the \enquote{signal} is supposed to be smooth, which cannot be said of most systems that undergo sharp transitions, for instance glacial terminations in paleoclimate data. Using a distance function readily applicable to raw or unevenly sampled data bypasses those issues. One aspect to note is that transforming from a pattern to a pdf in order to use the Wasserstein distance presumes that the physical processes underlying the time series are characterized by repeated modulations rather than recurrences of exact values. This might not be an adequate assumption for just any dataset.

\begin{figure}
    \includegraphics[width=\columnwidth]{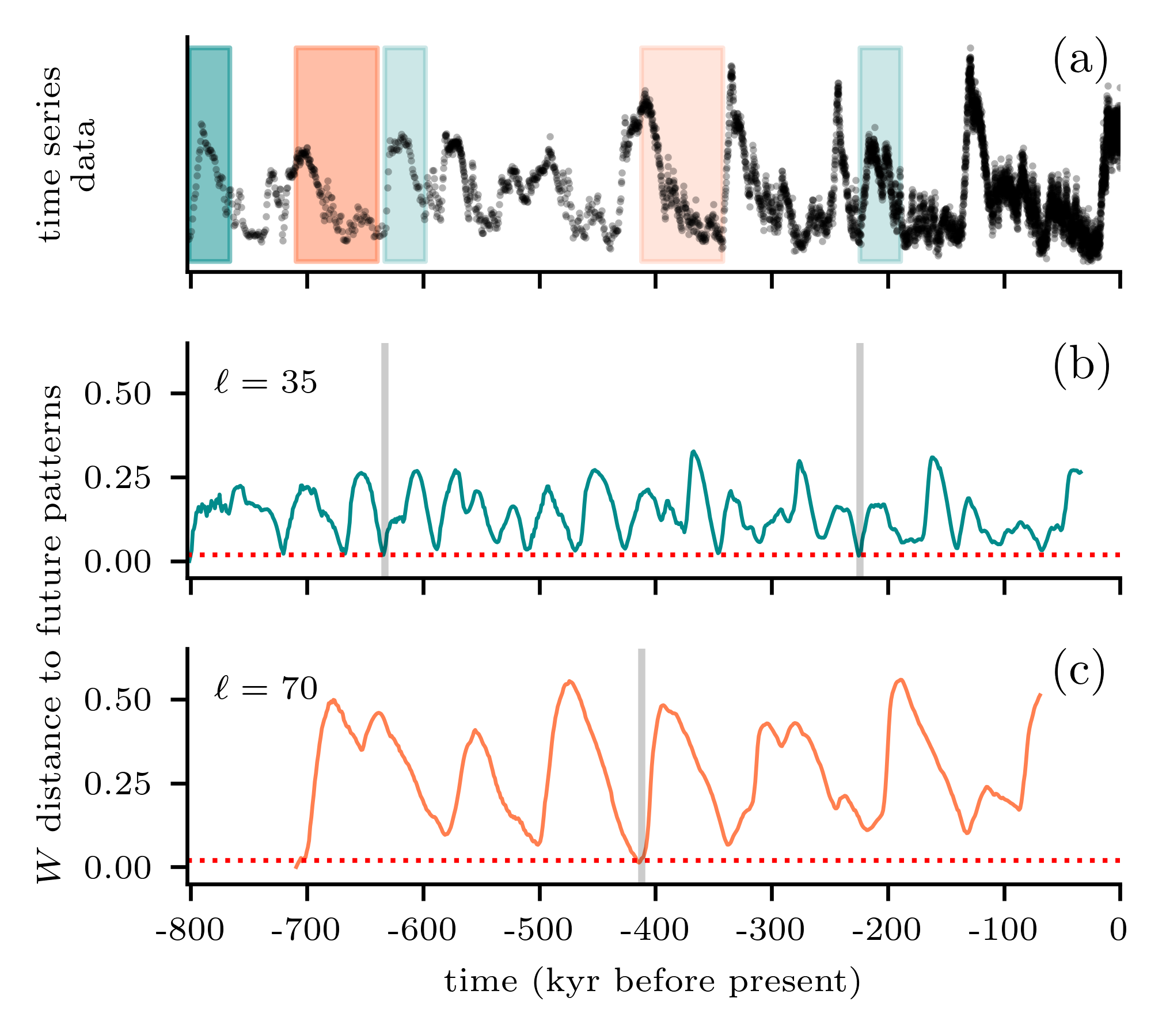}
    \vspace{-7mm}
    \caption{\label{fig:recurrence} Definition of pattern recurrences using local minima of the Wasserstein distance. (a) Two patterns (darker shaded areas on the left) of different lengths, \(\ell=35\) (blue) and \(\ell=70\) (orange), with \(\tau=1,000\) years or 1 kyr, are highlighted in an example time series. (b)-(c) Respective distance function to future patterns (blue and orange curves) with local minima characterizing recurrences (pale shaded areas in the top panel). The red horizontal dotted line represents the significance thresholds for \(p=0.01\) (see \textsection{\ref{sec:brownian}}) and vertical gray lines highlight the local minima that are below the significance threshold.}
\end{figure}

For any nonempty pattern \(x\in Y\), its distance to all future patterns in the time series forms a function of time as shown in Fig.~\ref{fig:recurrence}(b). Zeroes of this function would reflect exact recurrences of an identical event, up to a scaling constant. However, for most applications to real time series, the distance to all future patterns is likely to be strictly non-zero, meaning that any given pattern might occur exactly only once due to noise, imperfect data collection, or the nature of the dynamical system. Therefore, the definition of a recurrence is loosened using local minima of the distance function rather than zeroes. The local minima are defined with a search width in time of the order of the pattern duration \(\ell\tau\). This means that up until this point, our recurrence detection method is practically free of tunable parameters, with the only choices being the discrete time step \(\tau\) and recurrences duration \(\ell\).  

Yet, not all local minima are necessarily recurrences. Namely, even if a pattern does not recur, its distance to future patterns will still likely have local minima. For some applications, like qualitatively comparing climate epochs from different paleoclimate records, one could identify a few of the closest recurring patterns (\emph{i.e.} the global minimum as the most similar recurrence, and so on). More quantitatively, a recurrence can be defined as a local minimum of the distance to other patterns that is also below a certain "significance" threshold. Such a Wasserstein distance threshold for local minima to define a recurrence can be chosen based on percentiles of the distance distribution~\cite{Kraemer2018_Recurrence}. Depending on the context, the dataset and research questions can also inform the choice of local minima, as exemplified in \textsection\ref{sec:ird}, where we focus on local minima that happen at the same time for different time series. Alternatively, the following section introduces a principled Wasserstein distance threshold choice for recurrences in multiscale time series data. 

%

\section{Recurrences in Brownian motion \label{sec:brownian}}
\begin{figure*}
\includegraphics[width=\textwidth]{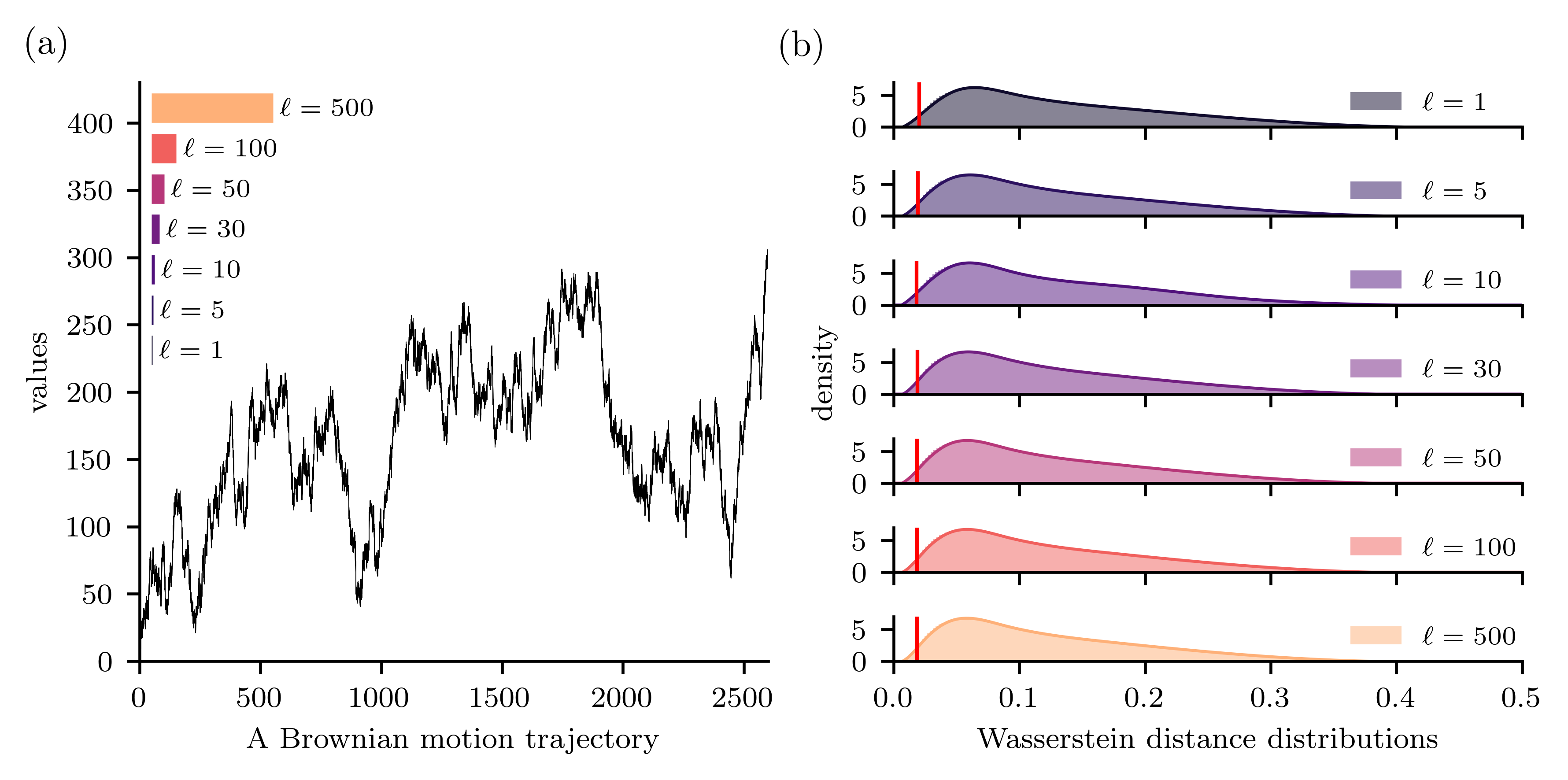}
\caption{\label{fig:null} (a) An example of Brownian motion trajectory and various pattern durations $\ell$. (b) Wasserstein distance empirical distribution for patterns in Brownian motion at different time scales. Each distribution represents the distance between the first pattern and subsequent non-overlapping patterns for many different trajectories such that we have a total sample size of $10^6$ pairwise distances for each duration \(\ell\in\{1,5,10,30,50,100,500\}\). Regardless of the time scale \(\ell\), the distances (0.019~\(\pm\)~0.001) and (0.032~\(\pm\)~0.002) correspond to p-values of 0.01 and 0.05, respectively highlighted by the red and black lines over the distributions and provided in Table~\ref{tab:thresholds}.}
\end{figure*}
In this section, we show that for a time-scale-invariant random process, the distribution of Wasserstein distances between patterns is also scale invariant, which provides a principled distance threshold choice for recurrences. We study the Wasserstein distance as defined by Eq.~\eqref{eq:my_wasserstein} between patterns generated with a numerical implementation of a one-dimensional random walk and find that the distance distributions are numerically invariant of pattern duration over two orders of magnitude. This invariance in distribution provides a statistical significance test for recurrences in multiscale time series, given that Brownian motion is an adequate null model for the time series under study. 

Brownian motion is one of the simplest stochastic processes. Also called the Wiener process in the one-dimensional case, it can be thought of as the continuous generalization of a random walk. It is defined as a stochastic process \(\{B(t)\,|\, 0\leq t<\infty\}\) on a probability triple \((\Omega, \mathcal{F}, \mathcal{P})\) with the following properties~\cite[\textsection 1]{Freedman1983_Brownian}
\begin{enumerate}
\item \(B(0, w)=0\) for each \(w\);
\item \(B(\cdot, w)\) is continuous for each \(w\);
\item for \(0<t_1<t_2<\hdots<t_{n-1}<t_n\), the increments
\[B(t_1), B(t_2)-B(t_1), \hdots, B(t_n)-B(t_{n-1})\]
are independent and normally distributed, with means 0 and variances \(t_1, t_2-t_1,\hdots, t_n-t_{n-1}.\)
\end{enumerate}
It follows from the latter property that it is scale invariant, \emph{i.e.} \(a^{-1}B(a^2 t)\) is also a Brownian motion for all \(a>0\). This property and its analytical simplicity make it an ideal null model for our applications, since it makes minimal assumptions about an underlying dynamical process -- only that increments are independent and normally distributed -- whilst generating time series that vary across multiple time scales. An example trajectory is shown in Fig.~\ref{fig:null}(a). 

Remarkably, the scale invariance property of Brownian motion reflects on the distribution of our Wasserstein distance implementation between patterns across time scales. Fig.~\ref{fig:null}(b) shows the empirical Wasserstein distance distributions between non-overlapping patterns for different recurrence time scales \(\ell\) spanning over two orders of magnitude. All distributions are based on a histogram of \(1\times10^6\) pairwise distances between non-overlapping patterns. More time series were sampled for longer pattern durations in order to achieve sample size. The Brownian motion trajectories were generated discretely using the cumulative sum of normal samples, with a resolution of 30 data points per time unit. Namely, this means that \(\ell=1\) and \(\ell=100\) patterns are respectively made of 30 and 3000 data points. Considering this vast range in resolution and time scale for all \(\ell\in\{1,5,10,30,50,100,500\}\), it is notable that all distributions seem to converge to the same probability density function. Most, if not all, other distance functions would have very different distributions across varying time scales, as illustrated with normalized and standard Euclidean distance in Fig.~\ref{fig:null_euclidean}, Appendix. This phenomenon has been an issue for time-delay embeddings and threshold choice in recurrence analysis~\cite{Kraemer2018_Recurrence}, especially at multiple time scales or when the temporal resolution varies over a time series. We observe a slight difference in the \(\ell=1\) curve, as also depicted in the cumulative distribution functions shown in Fig.~\ref{fig:null_cdf}, which we attribute to numerical sampling size effects\footnote{With fewer data points in the shortest \(\ell=1\) patterns, the number of possible time series patterns is reduced and the distribution does not converge exactly to the same invariant distribution as with longer patterns.}. Demonstrating the scale invariance analytically and obtaining closed-form expressions for the probability density functions of Wasserstein distance between patterns is still an open area of research. 

This invariance in distribution provides a mapping between any given value of \(W\) and the probability of encountering two patterns in Brownian motion at least as similar, regardless of their duration. This is the interpretation of the empirical cumulative distribution functions (cdf) illustrated in the Appendix, Fig.~\ref{fig:null_cdf}, and obtained from the pdf curves of Fig.~\ref{fig:null}. The cdf curves illustrate the probability of finding a non-overlapping pattern pair of equal duration in Brownian motion that has a given Wasserstein distance \(W\) smaller than given by the \(x\)-axis.

Under a given null model, a p-value is the probability of measuring an outcome at least as extreme as the one observed~\cite{Biau2010_Value}. Here, recurrences in Brownian motion provide a null model because recurring events can be detected as similar patterns without any underlying physical significance to this resemblance. The scale invariance of Wasserstein distance distributions between patterns in Brownian motion provides a principled distance threshold choice for recurrences in multiscale time series. This invariant distribution generates a mapping between the left p-value \(p\) and a distance threshold \(\varepsilon_p\), which has the following interpretation: if the distance between two patterns is less than \(\varepsilon_p\), then in a one-dimensional random walk there is a probability less than \(p\) to detect two patterns as similar.  The threshold is invariant of scale and independent of the time series under study, as it only depends on properties of patterns in Brownian motion.

The choice of Brownian motion as a recurrence null model is based on its simplicity, variability at many scales, its dynamical system nature (as opposed to decorrelated data points or a signal overlaid with random noise, for instance), and the empirical finding that the distance distribution of patterns does not vary with recurrence time scale \(\ell\). Other choices of null model could be considered; for instance, when dealing with particularly noisy data at only one time scale, one could work with the distance distribution in uncorrelated white noise of similar amplitude and sampling rate, such that the recurrence would be a local minimum that is \enquote{as unlikely as within purely uncorrelated noise}. Other threshold choice methods are available, for instance percentiles as in Ref.~\cite{Kraemer2018_Recurrence}, which are used in \textsection~\ref{sec:ird}. 

Some threshold values \(\varepsilon_p\) for a few standard \(p\) values for significant recurrences are provided in Table~\ref{tab:thresholds}. The mapping from a Wasserstein distance to the probability of such a distance arising between patterns in Brownian motion is available as a lookup table~\cite{code} and the cdf curves are plotted in Appendices, Fig.~\ref{fig:null_cdf}. As with all analyses involving statistical significance testing, rejecting the null hypothesis does not guarantee statistical significance~\cite{Nuzzo2014_Scientific}. Despite those limitations, in some contexts, this interpretation does provide a principled threshold choice for the Wasserstein distance between patterns, as exemplified in \textsection\ref{sec:edc}. The probability of encountering recurrences as similar in Brownian motion also offers the possibility to equip each recurrence with an error probability, which will be considered in future work.

\begin{table}
\begin{tabular}{c|c|c}
\hspace{0.3cm} P-value\hspace{0.3cm} & \hspace{0.3cm} Wasserstein distance \hspace{0.3cm} & \hspace{0.3cm} Error \hspace{0.3cm}\\\hline
0.01 & 0.019 & 0.001 \\
0.05 & 0.032 & 0.002 \\
0.10 & 0.041 & 0.003
\end{tabular}
\caption{\label{tab:thresholds}Wasserstein distance thresholds for a few standard p-values, considering a scale invariant Brownian motion null model for patterns recurrences. Error is 2$\sigma$ across time scales \(\ell\in\{1,5,10,30,50,100,500\}\).}
\end{table}

%

\section{\label{sec:network}Multiscale recurrence networks}

Recurring patterns as defined through the previous sections provide insights into the coarser temporal structure of complex time series. Here, we map them to a network, which acts as a data structure to facilitate information storage, navigation, and analysis. Time is divided into discrete time steps, which are nodes, before connecting pairs when they are both the initial time of a pattern recurrence. This yields a directed acyclic graph where the topological order is trivially dictated by the time ordering of the nodes. The duration of recurring patterns is stored in the network as an edge attribute. Repeated patterns in the time series are reflected in the existence of pairwise connections in time. Thus, the network structure is a tool to navigate temporal information embedded in time series recurrences at multiple time scales. 

\begin{figure}
\includegraphics[width=\columnwidth]{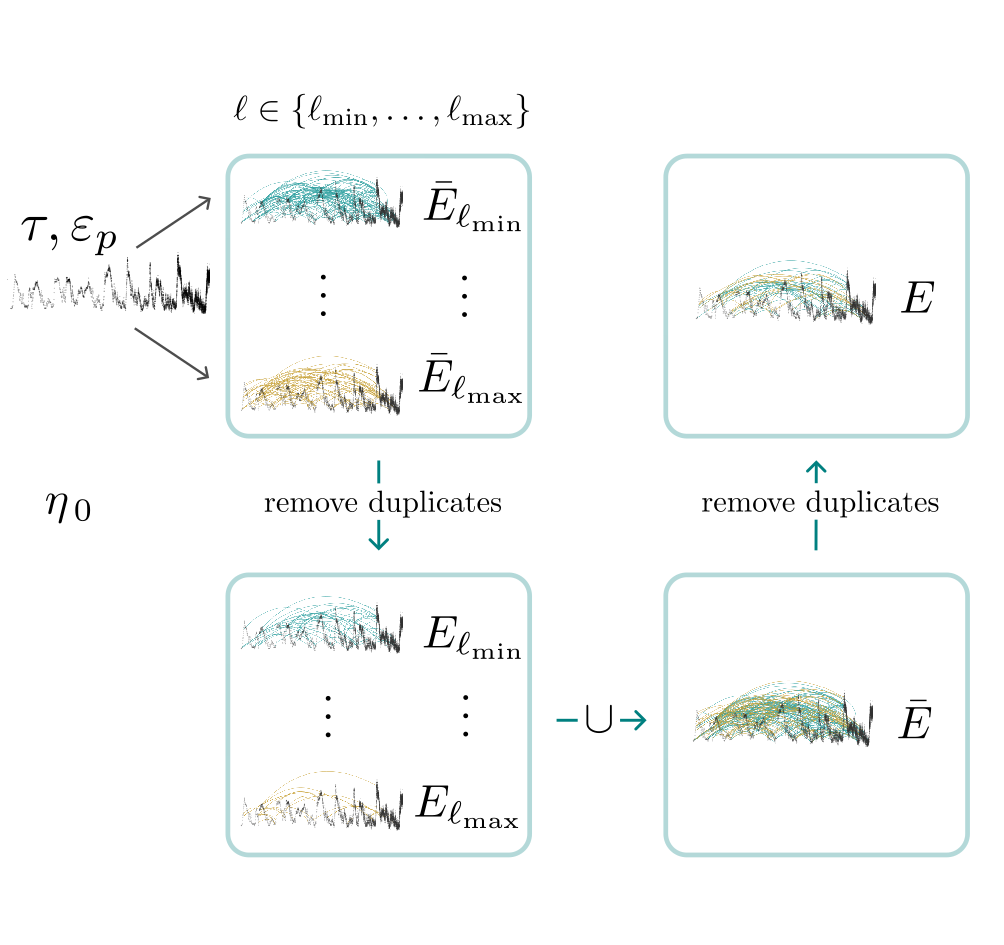}
\caption{\label{fig:pipeline}From detection at multiple time scales \(\ell\) to a multiscale recurrences set. A complete recurrence search is performed at every time scale \(\ell\in\{\ell_{\min},\hdots, \ell_{\max}\}\), using the same discrete time step~\(\tau\) and Wasserstein distance threshold~\(\varepsilon_p\). Duplicates are removed according to an overlap~\(\eta_0\), before combining recurrences across time scales and removing duplicates once more.}
\end{figure}

For a given time series, the ensemble of detected recurrences depends on the number of discrete time steps \(N\) and the time scale \(\ell\) at which the recurrence search is performed. First, a complete recurrence search is completed at every time scale \(\ell\), which yields separate edgelists over the same set of nodes. Then, those edgelists are merged into one multiscale edgelist, and overlapping recurrences are combined to avoid repetitions. All details of the method are provided below, which is also illustrated schematically in Fig.~\ref{fig:pipeline}. In what follows, we consider \(\tau=1\) to simplify the notation. 

For a given pattern of length \(\ell\) starting at time \(t_i\), a \emph{recurrence search} consists of computing the Wasserstein distance to all future patterns and identifying local minima below a significance threshold \(\varepsilon_p\). A recurrence \((i,j,\ell)\) is defined for such future events at time \(t_j\), where \(i\) and \(j\) are respectively the indices of recurring events at \(t_i\) and \(t_j\), and \(\ell\) is the duration. For instance, Fig.~\ref{fig:recurrence} shows two recurrences detected for the blue pattern and one for the orange pattern.

A \emph{complete recurrence search at time scale }\(\ell\) is performed when all future recurrences of all patterns of length \(\ell\) are detected. Starting from the oldest time step \(t_0\), a recurrence search is performed for recurrences of the \(x(t_0, \ell)\) pattern in all future time steps using the local minimum condition and significance threshold. If two recurrences are detected at both \((t_i, t_j, \ell)\) and \((t_{i+1}, t_{j+1},\ell)\), a single longer recurrence \((t_i, t_j, \ell+1)\) is defined. All the recurrences are stored in a \emph{complete edgelist at time scale \(\ell\)}, \(\bar{E}_\ell = [(i, j, \ell), \hdots]\). Recurrences in \(\bar{E}_\ell\) can still overlap with one another. If overlapping recurrences are not wanted, for instance to perform quantitative analyses based on recurrence counts, those can be combined using the following scheme. 

The standard measure of overlap between two intervals \(I = [a,b]\) and \(I'=[a', b']\) is given by
\begin{equation}\label{eq:overlap}
|I\cap I'| = \max[0, \min(b, b')-\max(a,a')],
\end{equation}
where we use standard set theory notation for the intersection and $|I|=b-a$ is an interval's length. In the context of multiscale recurrences, where the goal is to quantify the total overlap between two pairs of intervals, we define an overlap function \(\eta\) as follows
\begin{equation}\label{eq:my_overlap}
\eta(I, J, I', J') = \frac{|I\cap I'| + |J\cap J'|}{\ell+\ell'}
\end{equation}
where \(I\) and \(J\) are intervals of duration \(\ell\) respectively starting at \(t_i\) and \(t_j\) (and similarly for the primed recurrence). This yields a number between 0 and 1 that quantifies how much two recurrences capture the same intervals in the time series, with 1 only if both recurrences overlap perfectly and have the same duration. The purpose of using \(\ell+\ell'\) as a normalization instead of the maximum value of \(2\min(\ell, \ell')\) is to account for recurrences of different duration. With the normalization in Eq.~\ref{eq:my_overlap}, two recurrences that perfectly overlap but have widely different durations can still be considered distinct, depending on the overlap threshold chosen. 

For a given recurrence set \(\bar{E}_\ell\), one can compute the overlap \(\eta\) between all recurrence pairs and set a threshold \(\eta_0\) at which two recurrences are considered indistinct. In this work, we used \(\eta_0=0.8\), which means that recurrences that overlap more than 80\% are treated as duplicates. For every such set of overlapping recurrences, which effectively capture the same recurring event in the time series, we remove duplicates by combining them with one of the schemes listed in Table~\ref{tab:combine}. The simplest schemes are using the average starting times for each recurrence in the overlapping set and either the average or longest duration $\ell$. The \texttt{combine} scheme ensures the resulting recurrence covers all possible overlapping intervals. Practical annotated examples are provided as notebooks in the code~\cite{code}.

\begin{table}
\begin{tabular}{c|c|c|c}
Scheme & Output $t_i$ &  Output $t_j$ & Output $\ell$ \\\hline
\texttt{mean} & $\langle t_i \rangle$ & $\langle t_j \rangle$ & $\langle \ell \rangle$ \\
\texttt{longest} & $\langle t_i \rangle$ & $\langle t_j \rangle$ & $\max\{\ell\}$ \\
\texttt{combine} & $\min\{t_i\}$ & $\min\{t_j\}$ & \begin{tabular}{@{}c@{}}
  $\ell+\frac{1}{2}\big(\max\{t_i\}-\min\{t_i\}$\\
  \hspace{5mm}$+\max\{t_j\}-\min\{t_j\}\big)$ 
  \end{tabular}\\
\end{tabular}
\caption{\label{tab:combine} Possible schemes to combine overlapping recurrences depending on the physical interpretation of the underlying time series data. Maxima, minima, and arithmetic means $\langle\cdot\rangle$ are taken over the set of overlapping recurrences, often just a pair.}
\end{table}

A recurrence search is performed at every accessible time scale \((\ell_{\min}, \hdots, \ell_{\max})\), with the upper and lower bounds depending on the time series. For instance, the minimum temporal resolution for \(\ell_{\min}\) and the longest expected event for \(\ell_{\max}\). When all edgelists \(E_\ell\) are detected, one can study recurrences across time scales. Alternatively, edgelists can be combined into one multiscale edgelist \(\bar{E}\) containing all detected recurrences at any accessible time scales. This is akin to taking the set union of all edgelists. Depending on the desired independence of recurrences required for further analyses, one can combine overlapping recurrences once more in the resulting multiscale edgelist. Removing duplicates twice renders the final multiscale edgelist \(E\) more similar to a set union, where repetitions of the same set element appearing in more than one of the original sets are not preserved. With intervals, the definition of duplicate recurrences is not as strict as it would be with discrete elements but rather approximated by the overlapping function $\eta$ and an appropriate threshold and recurrence combining scheme.

Even if our approach builds on similar underlying ideas, it bears a few differences with existing recurrence network analysis methods, both in terms of the distance function used and the conditions that define a recurrence. Most other similarity-based methods rely on thresholding the Euclidean distance between subsets of the time series~\cite{Eckmann1987_Recurrence, Marwan2007_Recurrence, Silva2021_Time}, which is the principle of time-delay embeddings. In that case, the distance between patterns \((x,t)\) and \((x', t')\) is computed as if \(x\) and \(x'\) were Euclidean vectors, assuming an even temporal sampling. While successful in numerous theoretical applications, like reconstructing chaotic attractors and computing the invariant measure~\cite{Takens1981_Detecting}, using Euclidean distance on complex temporal data like financial, climatic, or biological time series has a few drawbacks. Takens's theorem applies to dynamical systems that are deterministic and evolving on a smooth manifold, which cannot be said of most real datasets. The Euclidean distance distribution changes over time scale and point density, which means that the threshold has to be adjusted for different pattern lengths and time series (see Appendix~\ref{sec:appendix}, Fig.~\ref{fig:null_euclidean}). The thresholding method based on the distribution of distances suggested in Ref.~\cite{Kraemer2018_Recurrence} differs from ours in that the threshold still depends on the distances within the time series under study, whereas ours is based on an external statistical significance test. Another drawback of existing methods, especially information-theoretic divergences, is again the need for even temporal sampling, which calls for interpolation and preprocessing, or otherwise using only a subset of the data points to perform the embedding and recurrence analysis~\cite{Donges2011_Identification}. The general family of network time series analysis methods is reviewed in Refs.~\cite{Silva2021_Time}, which offers more insight into the mechanics of each method and the relations between them. 

%

\section{\label{sec:palaeo_results}Applications for paleoclimate records}
\begin{figure*}
\includegraphics[width=\textwidth]{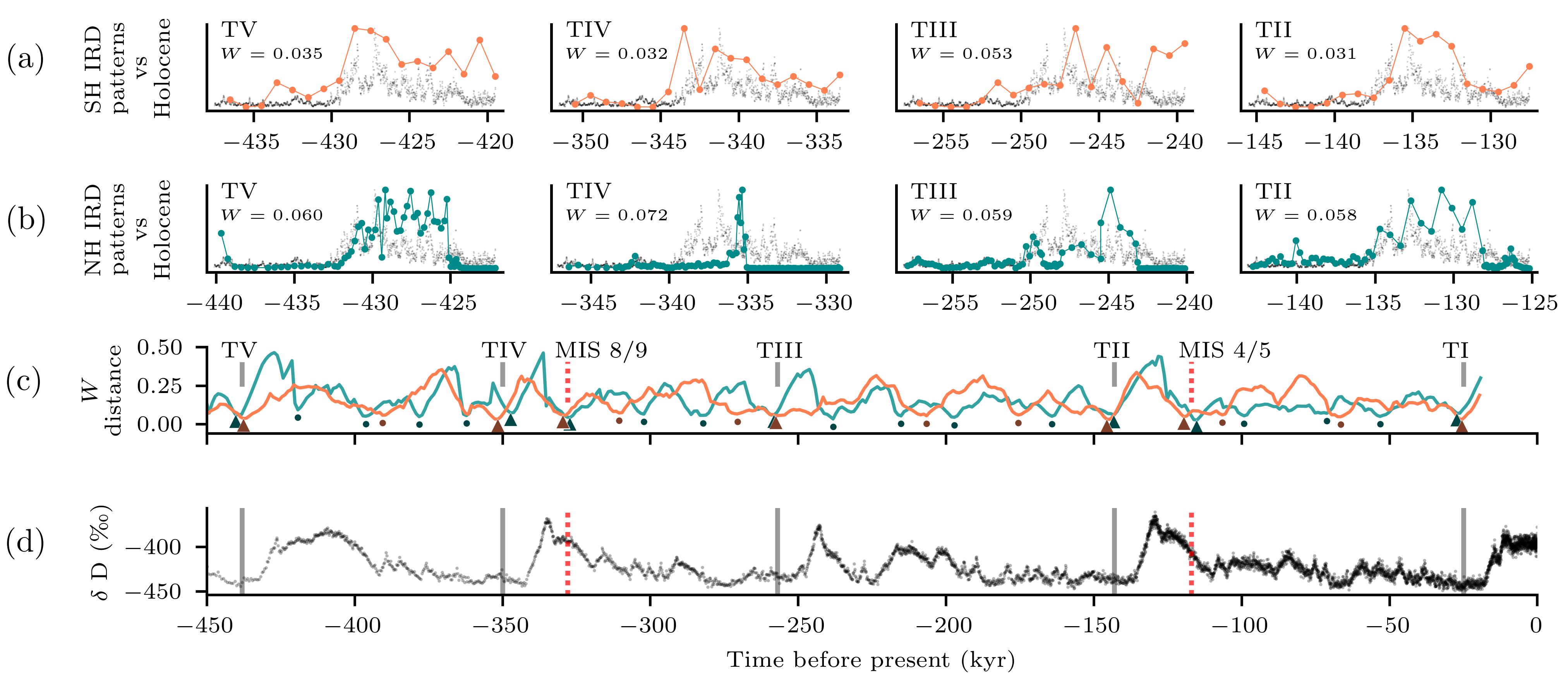}
\caption{\label{fig:ird} Identifying similar patterns in different IRD records. (a) Southern hemisphere (orange) and (b) Northern hemisphere (blue) IRD record excerpts, which align with terminations TII-V, overlaid with the Holocene record in pale grey for visual comparison. (c) Wasserstein distance function between Holocene and both the SH record (orange curve) and the NH record (blue curve). Local minima are identified underneath the curves by markers, with triangle markers highlighting local minima that are aligned in time for both hemispheres. (d) EPICA Dome C (EDC) ice core deuterium profile as a proxy for global temperature and showing glacial-interglacial periods. Vertical grey lines in (c)-(d) show terminations TI-V from Bajo \emph{et al.}~\cite{Bajo2020_Persistent} shifted by {$25-18=7$~kyr} to account for the start of the Holocene IRD record with respect to TI, and red dotted lines show additional local minima of the distance that align for both hemispheres during glacial inceptions of MIS 4/5 and MIS 8/9. Only panels (c) and (d) share the same x axis. }
\end{figure*}
In this section, we introduce two applications of our method on paleoclimate data, one where we find a specific pattern in different time series and another where all the recurring events at multiple time scales are extracted from a single time series.

\subsection{\label{sec:ird}Pattern similarity in ice-rafted debris data}

We provide a first example of how our method is applied to find ice-rafted debris (IRD) patterns akin to the Holocene (the interglacial period that we are currently in) throughout the Late Pleistocene. This latter period encompasses previous ice ages where both the Southern Hemisphere (SH) and Northern Hemisphere (NH) had extensive ice sheet cover. This example compares distinct records of analogous physical phenomena with different temporal resolution. 

Ice-rafted debris are larger-sized debris\footnote{Typically identified as sediment grains coarser than 2~mm in diameter.} found in marine sediment cores. These data are used as a proxy for ice sheet calving since only floating glacial ice can carry coarse debris from the ice sheet margin to sedimentation (and coring) sites. Cores retrieved from such sites are counted using X-ray scans or a sieving method, and their depth is transformed to an estimation of time using age models, which yields a record of IRD counts over time. We exemplify our method using IRD because these data are characterized both by sequences of zeros and by very abrupt shifts to high count layers, which are not easily treated in standard time series analysis methods and recurrence metrics. Furthermore, all three records have very unequal resolution. The Holocene stacked record from Weber~\emph{et al.}~\cite{Weber2014_Millennialscale} covers the period from 7 to 25 thousand years before present (ka BP), extracted from the Scotia Sea in Antarctica and with a decadal resolution of 5 to 15~years. Also from the Scotia Sea, the Southern Hemisphere record is published by Jasper~\emph{et al.}~\cite{Jasper2024_33MillionYear}. We use drilling site U1537, which covers the last 1.2 million years with a 1~kyr resolution. The Northern Hemisphere record is published by Barker~\emph{et al.}~\cite{Barker2019_Early}, covering the last 800~ka BP at a centennial resolution of 150 to 200~years, from Site 983 in the North Atlantic Ocean. 

As opposed to the global temperature proxy data shown in Fig.~\ref{fig:recurrence}, the IRD time series are not Brownian motion-like, and therefore the threshold choice proposed in \textsection\ref{sec:brownian} does not apply. Other methods can be used to choose the threshold under which local minima are considered recurrences. If only a few recurrences are desired to facilitate inspection and interpretation of the IRD patterns, one can manually choose the closest patterns by looking into the local minima with the smallest Wasserstein distances. To compare different records and identify all local minima below a threshold as recurrences, a percentile method as introduced in Ref.~\cite{Kraemer2018_Recurrence} is the approach we adopt here.

We compute the Wasserstein distances between the Holocene record and \(\ell=18\)~kyr long sliding windows every \(\tau=1\)~kyr of the last 450~ka BP of both the NH and SH IRD records. Local minima of the distance are detected within a search order of \(\ell\), and if they are below a Wasserstein distance threshold equivalent to the first quartile of distances (0.087 for SH and 0.095 for NH). The distance curves are shown on Fig.~\ref{fig:ird} (a), along with those local minima. Most patterns detected as similar to the Holocene record are happening at different times between the NH and SH records, which are highlighted by the round markers in Fig.~\ref{fig:ird} (c). However, there are 7 simultaneous SH-NH local minima to the Holocene (less than 4.5 kyr time difference between SH and NH local minimum), shown by triangle markers. As expected, there is a local minimum at the last glacial termination (TI) in both records, which reflects that IRD patterns of the Holocene are similar in all three records. 

All the Late Pleistocene major glacial terminations are captured by IRD patterns similar to the Holocene in both SH and NH records at the same time. The IRD count data corresponding to those four non-trivial similar patterns (terminations TII-TV) are plotted in Fig.~\ref{fig:ird} (a) and (b), along with the Holocene record for visual comparison and Wasserstein distance value at the local minima. Two other events, corresponding to Marine Isotope Stage (MIS) 5/4 and MIS 9/8 transitions from peak interglacial to glacial conditions, are found to be similar to the Holocene in both records at the same time. This reflects that ice sheet instability akin to the Holocene is present not only in major glacial terminations but also in periods of global cooling. 

In this example, the Wasserstein distance adequately reflects intuition of pattern similarity. For instance, the NH pattern closest to termination IV in Fig.~\ref{fig:ird}~(b) is visually quite different from the Holocene record. However, his dissimilarity is captured by the higher distance $W$, which could be excluded by another threshold choice. Another notable aspect is that both distance curves in Fig.~\ref{fig:ird}~(c) do not need to be shifted to be compared, despite the contrast in resolution for the NH and SH records. Our framework enables comparison of IRD patterns across records, despite different time resolution, age models, and complex data characterized by abrupt changes. 

\subsection{Antarctic ice core multiscale recurrence network\label{sec:edc}}

Our second example introduces a complete set of non-overlapping recurrences at multiple time scales for the longest available Antarctic ice core record. We cover two orders of magnitude of recurrence duration \(\ell\) with the same detection method and parameters. 

\begin{figure}
\includegraphics[width=\columnwidth]{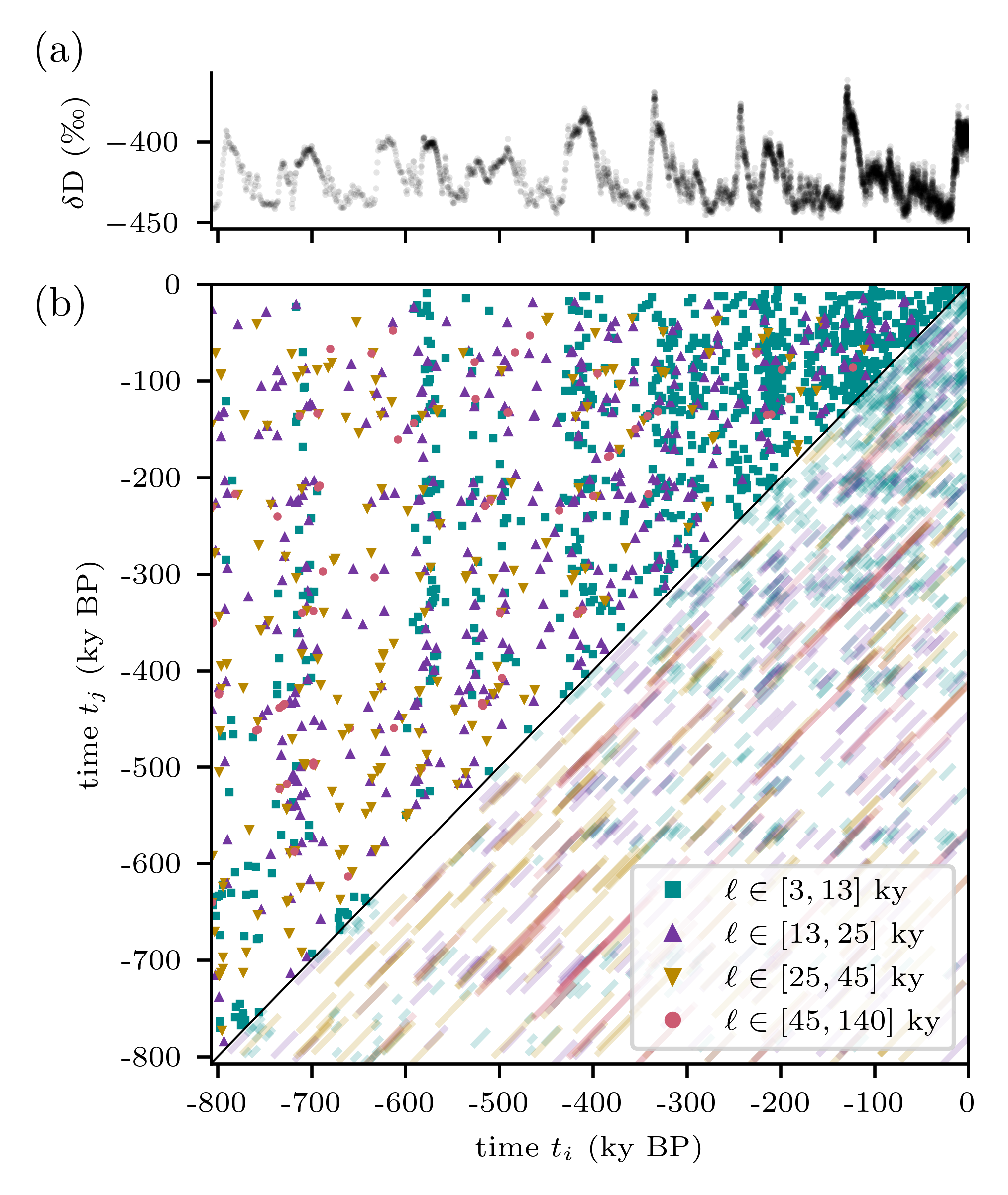}
\caption{\label{fig:EDC_matrix}Multiscale recurrence network for EDC ice core record. (a) EDC ice core deuterium ratio profile over time. (b) Matrix representation of the corresponding multiscale recurrence network. In the upper triangle, each marker illustrates one recurrence \((i,j,\ell)\) at starting times \((t_i, t_j)\), with color and marker shape given by recurrence duration \(\ell\). In the lower triangle, the same recurrences are plotted as traces from \((t_i, t_j)\) to \((t_{i+\ell}, t_{j+\ell})\).}
\end{figure}

The time series under scrutiny is the deuterium ratio profile in the European Project for Ice Coring in Antarctica Dome C (EDC) ice core~\cite{Jouzel2007_Orbital}, equipped with the most recent age model~\cite{Bouchet2023_Antarctic}. The deuterium ratio $\delta D$ in the EDC record is proportional to global temperature changes and captures high-resolution climate variability in the Late Pleistocene. The record is plotted in Fig.~\ref{fig:EDC_matrix}~(a), which shows the variability at multiple time scales, the notable change in resolution between older and newer parts of the record, and the large-scale oscillation between glacial and interglacial periods. We choose a discrete time step interval of \(\tau=1\)~kyr for a total of \(N=807\) nodes between \(t_0=-806.258\) and \(t_{N-1}=-0.258\)~ka BP. We perform recurrence detection for all event lengths \(\ell=3, 4, \hdots, 120\), with a Wasserstein distance threshold of 0.019 (see Table~\ref{tab:thresholds}). Recurrences that overlap more than 68\% are then combined, within and across time scales (\(\eta_0=0.68\), see Fig.~\ref{fig:overlap} and Appendix for additional details). The overlapping recurrences for which \(\eta<\eta_0\) are re-defined using the \texttt{combine} scheme from Table~\ref{tab:combine}. The lower bound \(\ell=3\) is chosen with the record's lowest temporal resolution of \(1.2\)~kyr in mind, such that recurring patterns are neither trivial (a single data point) nor restricted to higher-resolution parts of the record. 

The resulting set of climate recurrences comprises 1601 event pairs of durations between 3~kyr and 138~kyr. The complete set of recurrences is illustrated as a matrix in Fig.~\ref{fig:EDC_matrix}~(b). Each recurrence is plotted twice, as a marker in the upper triangular matrix and as a trace in the lower triangular matrix. The markers best represent the location of each individual recurrence in time, whereas the traces provide more insight into the duration and juxtaposition of recurrences across time scales. 

\begin{figure}
\includegraphics[width=\columnwidth]{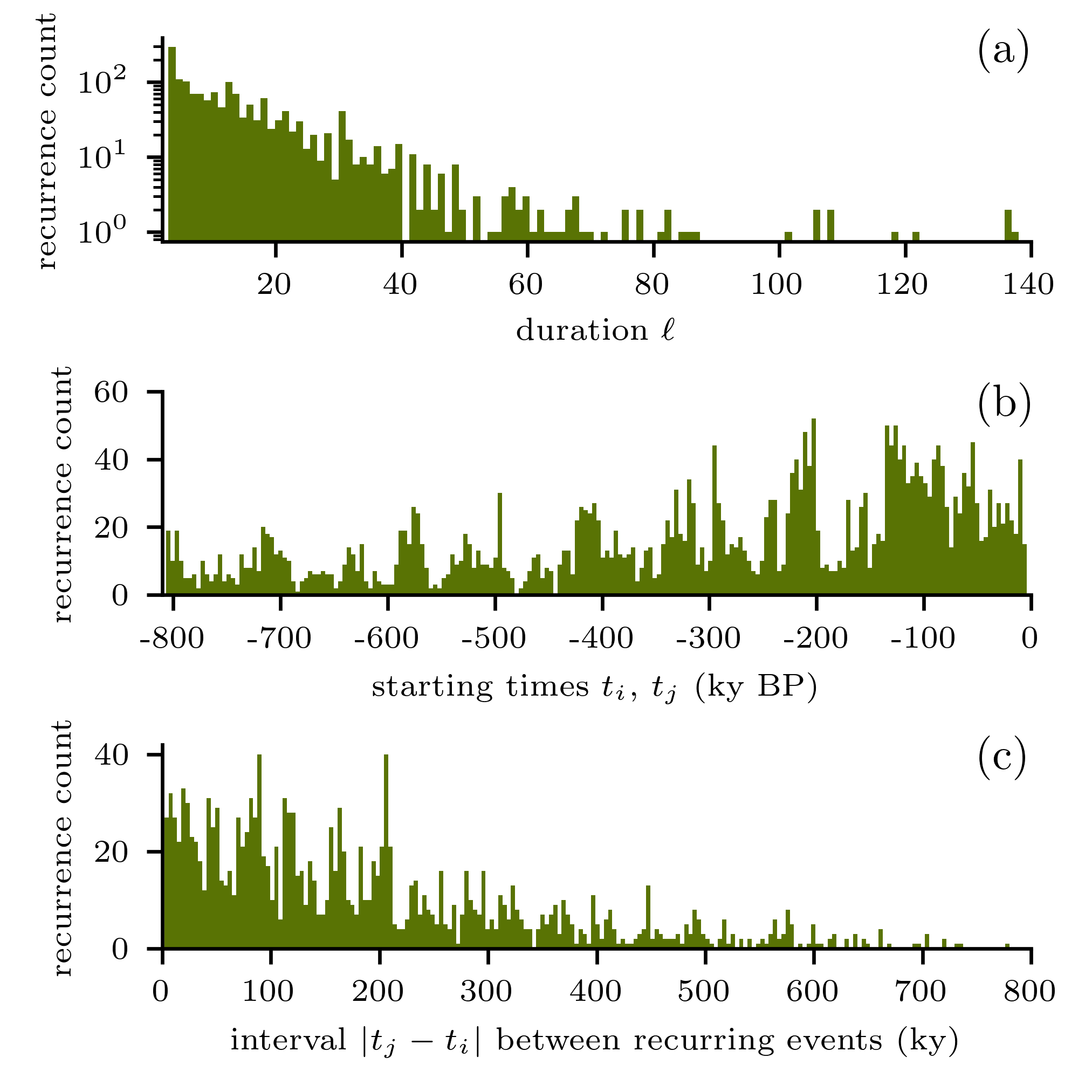}
\caption{\label{fig:EDC_results}Properties of the EDC multiscale recurrence network. (a) Distribution of durations \(\ell\) for all recurrences detected in the EDC ice core. (b) Distribution of starting times \(t_i, t_j\) for all recurrences \((i,j,\ell)\). (c) Distribution of intervals \(|t_j-t_i|\) between an event at time \(t_i\) and its recurrence at time \(t_j\). }
\end{figure}

Figure~\ref{fig:EDC_results} characterizes some properties of the EDC multiscale recurrence network. Durations \(\ell\) in panel (a) follow a long-tailed distribution, which highlights that short-scale events are more numerous and diverse, but longer recurrences up to \(\ell=138\)~kyr capture glacial/interglacial larger climate oscillations. Fig.~\ref{fig:EDC_results}~(b) shows where recurrences are located in time. This panel is a projection of the matrix in Fig.~\ref{fig:EDC_matrix} on the time axis. Both of these plots show more numerous connections between abrupt climate events like glacial terminations and inceptions but also reveal connections throughout the record between these and shorter events within and across glacial cycles. The distribution of time between events across time scales reveals a persistent pacing of the Late Pleistocene, which is explored further in separate work, along with a more detailed climatic analysis of the EDC multiscale recurrence network. 

\section{Conclusion}

We have introduced a framework to use the well-known Wasserstein metric to detect recurring patterns in messy or unevenly sampled time series data. Using our method, similarity can be quantified regardless of temporal sampling while exactly preserving  patterns' shape, the duration of events and the richness of the underlying data. Recurrences are defined as local minima of the distance and an additional threshold can be used depending on the context and research questions. Our approach supports meaningful detection of similar patterns in challenging records or across different time series. 

When applied to Brownian motion trajectories, our definition of the distance between patterns is numerically invariant in distribution across more than two orders of magnitude of patterns' duration or time scale. This invariance provides a mapping between the Wasserstein distance and the probability of pattern pairs being found as similar in a random walk. For a multiscale time series, which operates at many different time scales all at once, Brownian motion can be an adequate null model. In this case, there is a principled threshold choice for local minima of the distance to be considered significant recurrences. We provide a table of Wasserstein distance thresholds depending on the desired p-value of pattern similarity compared to a random walk null model. 

We then introduced a multiscale recurrence network framework to study the temporal structure of recurring patterns in time series across multiple time scales. One can either study the ensemble of recurrence networks at different time scales or combine them into one larger multiscale network. We introduced a measure of overlap between two recurrences, such that recurring events associated with each edge can be considered distinct up to a chosen level of overlap. The multiscale network encodes the information of moments in time connected by similar patterns in the time series for various pattern durations. Compared to existing methods, our approach allows quantitative analyses on the recurrence counts over time, the time intervals between an event and its recurrence, and the interplay between different time scales or durations. 

Finally, we illustrated our framework on two different paleoclimate applications. Our novel recurrence detection method is ideal for uneven temporal resolution or highly non-continuous data like ice-rafted debris in marine sediment cores. As an example of finding similar patterns across different records, we detect IRD patterns similar to the Holocene (last 25 ka BP) in longer marine sediment records from the Northern and Southern hemispheres. We then introduced the multiscale recurrence network of the EPICA Dome C ice core, a record of global temperature variability in the last 800 ka BP, with recurrences spanning over two orders of magnitude in duration across multiple glacial cycles.

Characterizing recurring events in complex time series is only a first step towards a broader understanding of the physical phenomena that generated those recurrences. To further explore the underlying causes of recurring patterns, one could use a causal inference framework~\cite{Runge2023_Causal} or transfer entropy for longer recurrences and denser time series~\cite{Kirkley2025_Transfer}. In follow-up work on the EDC ice core recurrence network, we explore the relation between the timing of predominant recurrences and that of external orbital forcings, which influence global climate patterns. An interesting application of the scale invariance of our recurrence detection framework will be to study the fractal structure of various time series, including the climate, but also biological and economic time series. Our approach is targeted to understanding the past rather than predictive applications, for which transformer-based methods are a promising avenue to integrate multiscale and multivariate information~\cite{Naghashi2025_multiscale}.

\begin{acknowledgments}
This work was supported by the Natural Sciences and Engineering Research Council of Canada, Te Herenga Waka Victoria University of Wellington, and the Antarctic Research Centre at Te Herenga Waka Victoria University of Wellington, Aotearoa New Zealand. NRG is supported by grants ANTA1801 (“New Zealand Antarctic Science Platform”) and RTVU2206 (“Our Changing Coast”) from the
New Zealand Ministry of Business, Innovation, \& Employment.
\end{acknowledgments}

\section*{Code availability}
Code implementing analyses and figures is written in Python and can be found at~\href{https://github.com/bdesy/wasserstein_recurrences}{\texttt{https://github.com/\linebreak bdesy/wasserstein\_recurrences}}, along with instructions on where to access the paleoclimate archives explored in this work, all publicly available.

\section*{Author contribution}
B.D., M.L.R and N.R.G. designed the research, as well as analyzed and interpreted the results. B.D. wrote the codes for the numerical experiments and analyses. B.D. performed the research and wrote the manuscript. H.I. contributed to the design and analysis of \textsection\ref{sec:ird}. All authors read, commented, edited, and approved the final version of the manuscript.\\

\newpage
\appendix*

\section{Additional figures\label{sec:appendix}}

\begin{figure}[htb!]
\includegraphics[width=\columnwidth]{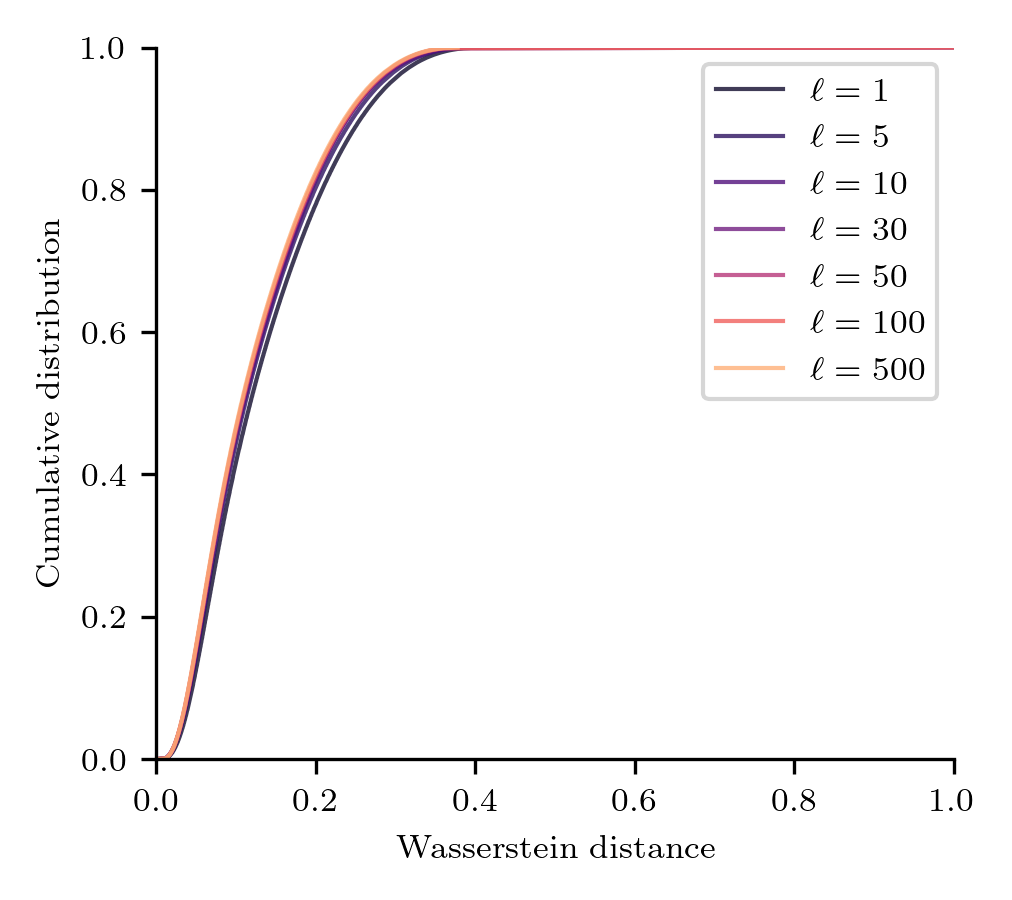}
\caption{\label{fig:null_cdf} Cumulative distribution functions of the Wasserstein distance between patterns in Brownian motion at different time scales, obtained by numerically integrating pdf from Fig.~\ref{fig:null}. The pdf are estimated using a smooth spline interpolation of degree 3 over the histograms, then converting negative values (which are artifacts of the spline interpolation) to zero and re-normalizing. }
\end{figure}

\begin{figure*}[htb!]
\includegraphics[width=\textwidth]{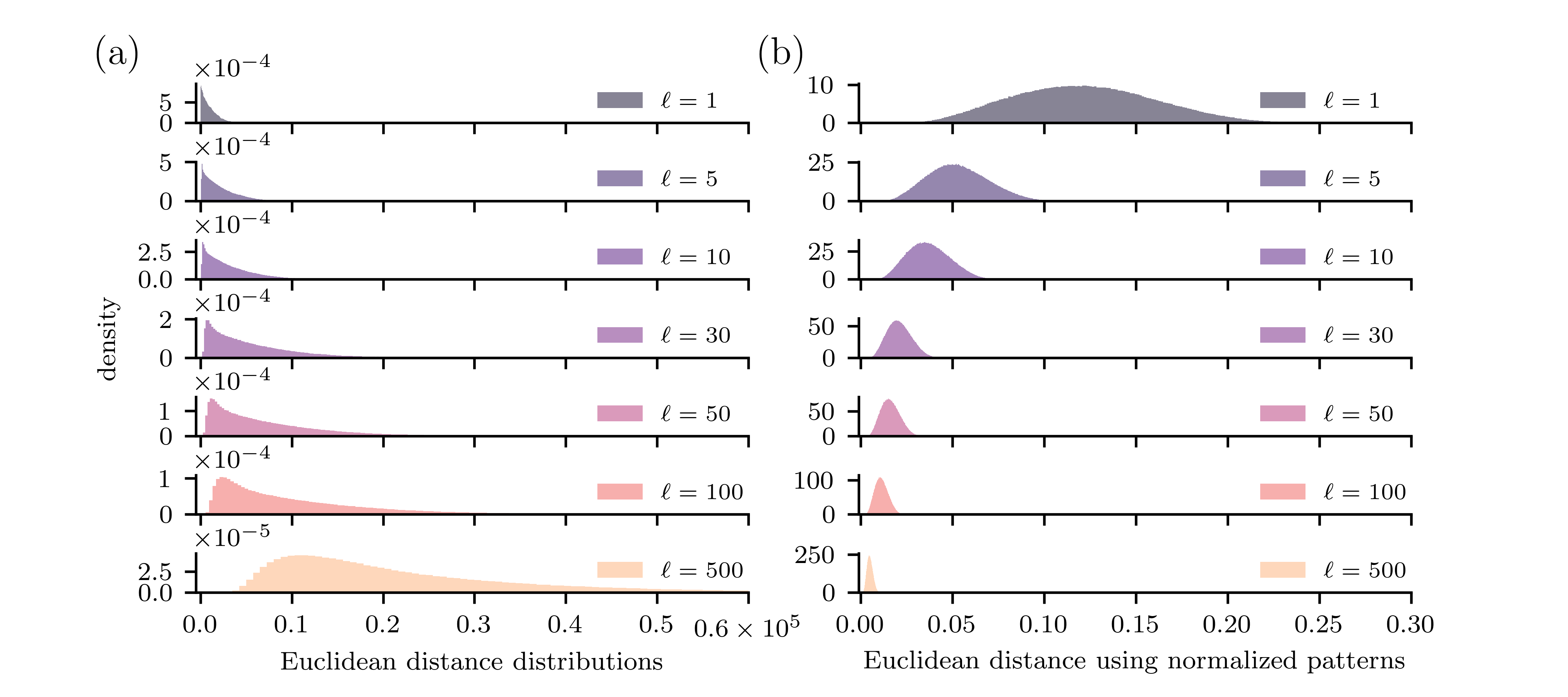}
\caption{\label{fig:null_euclidean} (a) Euclidean distance distributions between patterns in Brownian motion at different time scales. (b) Euclidean distance distributions between patterns in Brownian motion when patterns are normalised using Eqs.~(\ref{eq:discrete_measure}-\ref{eq:lower_bound}) before computing the standard Euclidean distance. In both cases, the distributions shift rather dramatically with time scale.}
\end{figure*}

\begin{figure}[htb!]
\includegraphics[width=\columnwidth]{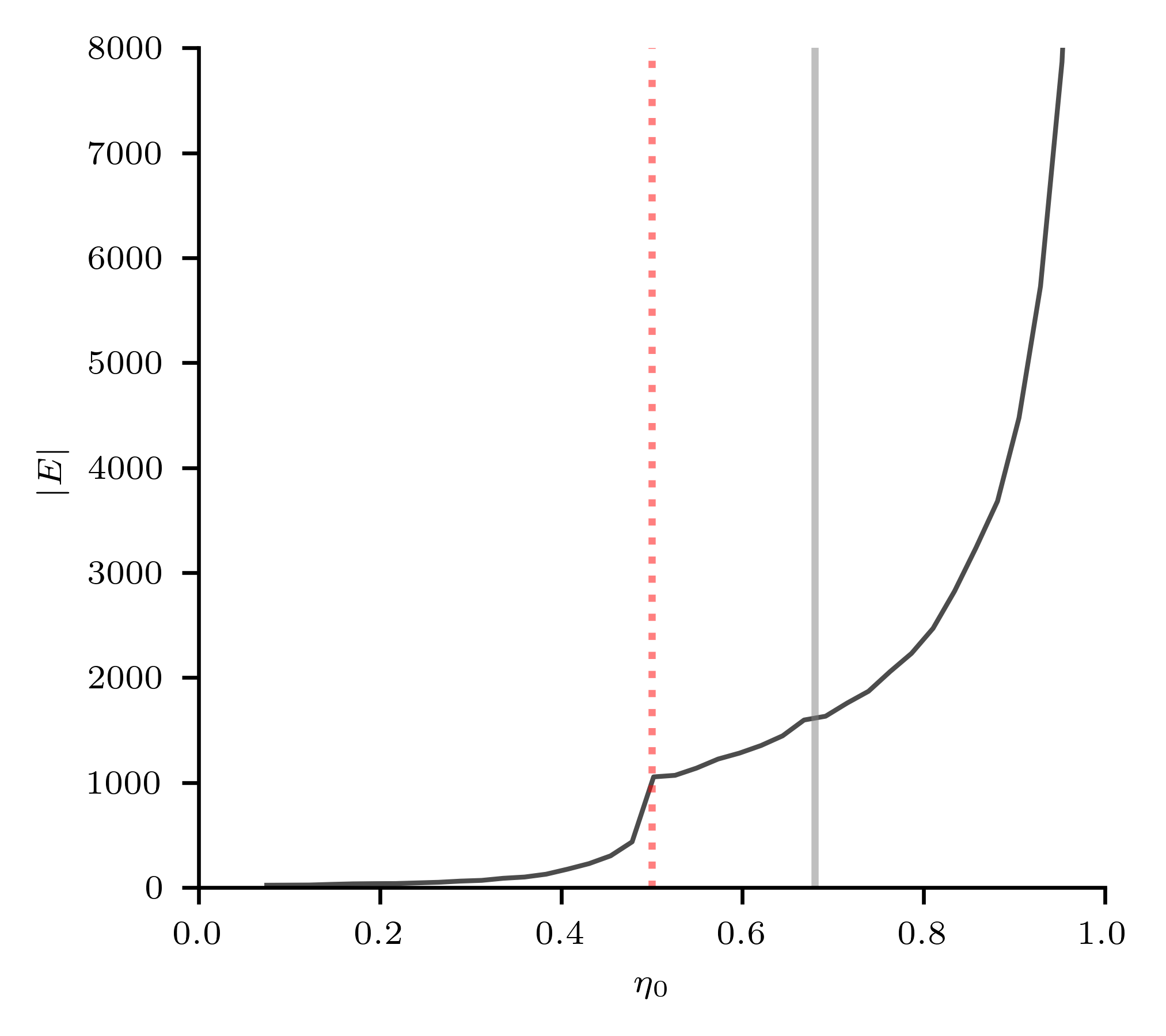}
\caption{\label{fig:overlap} Number of edges in EDC multiscale recurrence network with increasing overlap threshold $\eta_0$. The step at \(\eta_0=0.5\) and the chosen \(\eta_0=0.68\) for the EDC analysis are respectively highlighted by the dotted red line and the gray line.}
\end{figure}

In Fig.~\ref{fig:overlap}, the size of the EDC multiscale recurrence network shows two distinct asymptotic behaviors with overlap threshold parameter \(\eta_0\). For \(\eta_0\ll 0.5\), any recurrences with a slight overlap are combined, hence the network size collapses to a very few recurrences. For \(\eta_0\gg 0.5\), the number of recurrences explodes, because a very high number of duplicates are detected within and across time scales. There is a step-like transition at \(\eta_0= 0.5\), which can be explored using the definition of \(\eta\) in Eq.~\ref{eq:my_overlap}. Let us consider \(\ell=\ell'\), such that 
\begin{equation}\label{eq:explain_eta05}
\eta = \frac{1}{2}\bigg(\frac{|I\cap I'|+|J\cap J'|}{\ell}\bigg).
\end{equation}
The value \(\eta_0=0.5\) corresponds to a total overlap between both event pairs of the order of the recurrence duration \(\ell\). Further work with more records and synthetic time series would be needed to assess if this step-like behavior is universal or proper to this record. We hypothesize that for this particular system captured by the EDC ice core deuterium record, the total number of distinct recurrences is detected at \(\eta_0=0.68\). For this value, we exclude the two asymptotic behaviors and the gradient of the curve in Fig.~\ref{fig:overlap} is minimal, which means a small perturbation of \(\eta_0\) would have less impact on the number of recurrences detected.

\bibliography{biblio}

\end{document}